\documentclass[fleqn,usenatbib]{mnras}

\usepackage{newtxtext,newtxmath}

\usepackage[T1]{fontenc}

\DeclareRobustCommand{\VAN}[3]{#2}
\let\VANthebibliography\thebibliography
\def\thebibliography{\DeclareRobustCommand{\VAN}[3]{##3}\VANthebibliography}

\usepackage{graphicx}	
\usepackage{amsmath}	

\title[Clumpy model for V5668 Sgr]{A two-component clumpy model for the shell evolution of classical novae: the case of V5668 Sgr}

\author[Z. Abraham et al.]{
Zulema Abraham,$^{1}$\thanks{E-mail: zulema.abraham@iag.usp.br}
Larissa Takeda,$^{1}$
Pedro P. B. Beaklini,$^{2}$
Marcos Diaz,$^{1}$  
Kim L. Page$^{3}$
\newauthor
Laura Chomiuk$^{4}$
and
Justin D. Linford$^{2}$
\\
$^{1}$Instituto de Astronomia, Geof\'isica e Ci\^encias Atmosf\'ericas, Universidade de S\~ao Paulo \\
Rua do Mat\~ao 1226, CEP 05508-090, S\~ao Paulo, Brazil\\
$^{2}$National Radio Astronomy Observatory, 1003 Lopezville Road, Socorro, NM 87801, USA\\
$^{3}$School of Physics and Astronomy, University of Leicester, University Road, Leicester LE1 7RH, UK \\
$^{4}$Center for Data Intensive and Time Domain Astronomy, Department of Physics and Astronomy, Michigan State University,
East Lansing, MI 48824, USA\\
}

\date{Accepted XXX. Received YYY; in original form ZZZ}

\pubyear{2015}

\begin{document}
\label{firstpage}
\pagerange{\pageref{firstpage}--\pageref{lastpage}}
\maketitle

\begin{abstract}
 The shell of the classical nova V5668 Sgr was resolved by ALMA at the frequency of 230 GHz 927 days after eruption, showing that most of the continuum bremsstrahlung emission originates in clumps with diameter smaller than $10^{15}$ cm.  Using VLA radio observations, obtained between days 2 and 1744 after eruption, at frequencies between 1 and 35 GHz, we modeled the nova spectra, assuming first that the shell is formed by a fixed number of identical clumps, and afterwards with the clumps having a power law distribution of sizes, and were able to obtain the clump's physical parameters (radius, density and temperature). We found that the density of the clumps decreases linearly with the increase of the shell's volume, which is compatible with the existence of a second media, hotter and thinner, in pressure equilibrium with the clumps. We show that this thinner media could be responsible for the emission of the hard X-rays observed at the early times of the nova eruption, and that the clump's temperature evolution follows that of the super-soft X-ray luminosity.
We propose that the clumps were formed in the radiative shock produced by the collision of the fast wind of the white dwarf after eruption, with the slower velocity of the thermonuclear ejecta. From the total mass of the clumps, the observed expansion velocity and thermonuclear explosion models, we obtained an approximate value of 1.25 M$_{\odot}$ for the mass of the white dwarf, a central temperature of $10^7$ K and an accretion rate from   the secondary star of $10^{-9}-10^{-8}$ M$_{\odot}$ y$^{-1}$.
\end{abstract}

\begin{keywords}
binaries:close -- novae -- radio continuum: individual (V5668 Sgr) -- X-rays: individual (V5668 Sgr)
\end{keywords}



\section{Introduction}
\label{Introduction}
Multi wavelengths observations of nova eruptions during the last decade helped draw a clearer picture of their subsequent evolution \citep*[see review by][] {chom21}. 
The eruption, observed initially at optical wavelengths, is produced by a thermonuclear runaway in the surface of a white dwarf in a close binary system, which occurs when the physical conditions at the base of the accreting envelope reach a critical value  \citep*{sta72}.
During the event, $(10^{-7}-10^{-4})$ M$_\odot$ are ejected with velocities of  $\sim (500-3000)$ km s$^{-1}$, although in some cases velocities up to 10,000 km s$^{-1}$ were observed.
As the ejecta expands,  the brightness at optical wavelengths increases reaching a first maximum, decreasing afterwards, as the optical depth decreases. During the decay, some novae present further oscillation's or jitters in their light curve  \citep{str10}.
Studies of novae for which high resolution spectroscopy is available, showed that all of them present a slow P Cygni component before the first visual maximum. After the optical peak, new faster components can be seen in the spectra, superimposed to the slower P Cygni profile, indicating the existence of simultaneous flows  \citep{ayd20a}.

\begin{figure*}
\includegraphics[width=17cm]{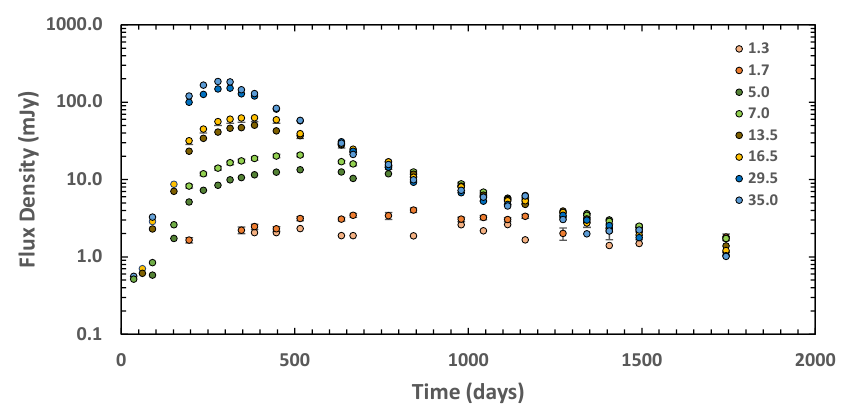}
	\caption{Light curves of V5668 Sgr  obtained with the VLA at the frequencies of 1.3, 1.7, 5.0, 7.0, 13.5, 16.5, 29.5 and 35.0 GHz. The time is measured from the beginning of the nova eruption (2015 March 15).}
    \label{fig:Fig_1}
\end{figure*}

The faster winds are probably produced by steady nuclear reactions in the surface of the white dwarf; when they overtake the slower material ejected during the eruption, they produce radiative shocks \citep{met14}, which are revealed by the observation of high energy (>100 MeV) $\gamma$ rays \citep{ack14}, emitted by particles accelerated to relativistic energies at the shock site. 
 An excellent correlation between $\gamma$-ray and optical light curves was found for the small number of novae for which simultaneous data are available, which shows that the optical emission after the first maximum is  produced both at the surface of the white dwarf and by material in the shock \citep{li17,ayd20a}.
Sometimes, early time radio emission is also correlated with $\gamma$-rays and it is attributed to non-thermal synchrotron emission, but it is not present in all radio light curves  \citep{cho21}.
As the shocked shell expands, it becomes transparent to high energy photons and eventually super-soft X-rays emitted at the surface of the white dwarf are observed  \citep{sch11}; they ionize the ejecta that produce thermal radio emission.
Also, hard X-rays (> 1 keV) are detected, emitted by the high temperature $(\sim 10^7$ K) gas from the shocked region  \citep{muk01}.

Radio light curves were used to obtain physical parameters of the expanding nova shell, employing different models \citep[see review by][]{roy12}. 
All the models assumed a homogeneous spherical shell with a power law density distribution: 
$n(r)\propto r^{-p}$, where $r$ is the radius of the shell  and  $p\sim 2-3$. 
In the "Hubble-flow" model, the mass ejection is instantaneous, the same amount of gas is expelled at all velocities, so that $p=2$, and a radial linear velocity gradient is assumed, with the fastest expanding material on the outside.
The variable wind model assumes a constant mass loss over a long period of time, so that the shell becomes a spherical region without a central cavity. 
Since eventually the wind ceases, a combination of the two models is generally assumed \citep{hje96}. 
The models reproduced the general shape of the light curves but obtained a mass for the ejecta that was up to one order of magnitude larger than that predicted by models based on the  relation between the mass of the ejecta and the decaying time of the bolometric luminosity  \citep{yar05,roy12,wen17}.

High angular resolution observations of  nova V5668Sgr, obtained in 2016 with ALMA at the frequency of 230 GHz showed, for the first time, that the millimetre wave emission arises from compact clumps distributed along the expanding shell. 
Recently,  multifrequency light curves of this nova obtained with the  Karl G. Jansky Very Large Array (VLA) at frequencies ranging between 1.3 and 35 GHz also became available \citep{cho21}.

In this paper we present a model in which we assume that the clumps detected with ALMA were formed early in the nova evolution, due to the turbulence that arises during  shock formation.  In our model we assumed that the expanding shell contains a fixed number of clumps,  which expand at constant temperature (obtained from optical models) as they move away from the white dwarf with constant velocity, and determined their density evolution by fitting their bremsstrahlung spectra to the observations. 
We found that the expansion of the clumps as a function of time is compatible with the assumption that they are in pressure equilibrium with a lower density and higher temperature medium, which is responsible for the X-ray emission observed at the beginning of the nova evolution.

In Section 2 we describe all the information available for nova V5668 Sgr, in Section 3, the observations used in this paper, in Section 4 our model and the  spectral fitting, in Section 5 we discuss the results and in Section 6 we present our conclusions.
\section{V5668 Sgr}

V5668 Sgr is a Fe II class classical nova, it  was discovered on 2015 March
15 \citep{sea15} and was followed up at optical and infrared wavelengths \citep{ban16,taj16,geh18,tak21}, as well as at $\gamma$ and X-ray energies \citep{che16,gor21},  and mm-wave and radio frequencies \citep{dia18,cho21}.
The first maximum in the optical light curve of V5668 Sgr occurred 6 days after the eruption; afterwards the luminosity remained almost constant for about 80 days, with several small amplitude quasi periodic oscillations \citep{ban16,taj16}. A high resolution optical spectrum was obtained 69 days after eruption \citep{taj16}; it shows broad emission lines together with blueshifted absorption lines: one at 768 km s$^{-1}$ and six with velocities between 1400 and 2200 km s$^{-1}$, representing multiple or anisotropic shell ejections. Although this was the first high resolution spectrum available for V5668 Sgr, spectra of other novae obtained earlier in their evolution show that only the low velocity absorption line is present before the first maximum in the optical light curve, while the higher velocity lines appear afterwards \citep{ayd20b}.
A strong decrease in the optical brightness  was observed 80 days after eruption, consistent with dust formation and confirmed by infrared observations \citep{ban16,geh18}.
High energy $\gamma$-rays were detected two days after the first optical maximum (8 days after eruption), followed by sporadic detections lasting for 53 days \citep{che16,gor21}. Although a correlation between the optical light curve and $\gamma$-rays with energies larger than 100 MeV could not be obtained due to the low $\gamma$-ray luminosity and exposure time, this correlation was found in nova V5856 Sgr \citep{li17}, implying that part of the optical emission after the first maximum arises from re-processed shocked material and not only from the white dwarf itself \citep{met14,met15}.
X-rays from V5668 Sgr, with energies $(0.3-10)$ keV were  first detected by the XRT telescope of the Neil Gehrels Swift Observatory on day 95, although observations had already started the second day after eruption,  with negative results \citep{geh18}. The X-ray spectrum presented both  hard $(1 - 10)$ keV and  super soft $(0.3 - 1.0)$ keV components; they
are attributed to emission from shocked material and from the surface of the white dwarf, respectively. The soft component rose by 4 orders of magnitude between days 100 and 200, presenting short term variability in which the flux dimmed and brightened again in a two or three days interval. The maximum in the soft X-rays light curve occurred on day 224;  the source had faded by around a factor of 1000 by the time of the last observation on day 848 \citep{geh18}.

\begin{figure}
\includegraphics[width=\columnwidth]{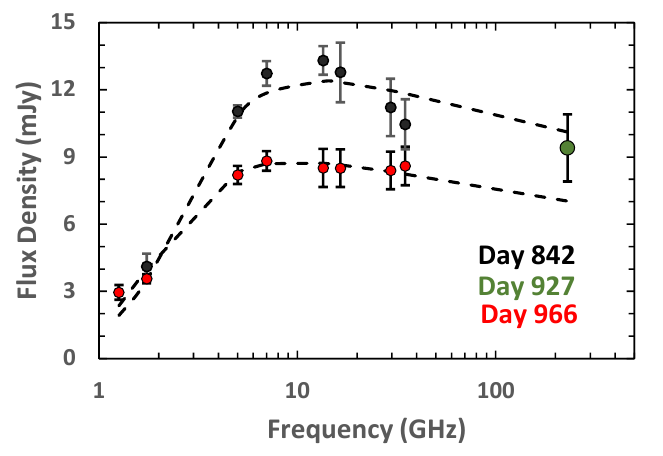}
	\caption{Spectra of V5668 Sgr obtained with the VLA on days 842 (black
circles) and 966 (red circles) after eruption, and flux density at 230 GHz
(large green circle) measured with ALMA on day 927. The broken lines represent
the result of the model described in Section 4.}
    \label{fig:Fig_2}
\end{figure}

\begin{table*}
    \centering
     \caption{X-ray fluxes and electron temperatures of nova V5668 Sgr}
    \begin{tabular}{c|c|c|c|c|c|c|c|c|c}
    \hline
    $t$  &  $t_{\rm dur}$ & & $L_X$ & $-\Delta L_X$ &  $+\Delta L_X$ & $T_X$ & $-\Delta T_X$ & $+\Delta T_X$ \\
    days after eruption
&  days & & $10^{32}$ ergs s$^{-1}$ & $10^{32}$ ergs  s$^{-1}$ & $10^{32}$ ergs s$^{-1}$ & $10^6$ K & $10^6$ K & $10^6$ K\\
    \hline
    100.9 & 5.56 & & 1.05 & 0.20 & 0.17 & > 314.5\\
    112.1 & 2.10 & & 1.25 & 0.30 & 0.38 & >  54.5\\
    117.6 & 1.74 & & 1.71 & 0.57 & 0.43 & > 78.9\\
    129.7 & 3.06 & & 1.79 & 0.38 & 0.43 & 100.0 & 47.6 & 334.8\\
    143.4 & 3.10 & & 3.44&  0.32&  0.85 & > 76.6\\
    171.0 & 0.01 & & 11.7 & 2.19 & 3.55 & 79.2 & 34.4 & 254.0\\
    186.3 & 0.01 & & 22.4 & 3.24 & 2.38 & 31.4 & 5.9 & 7.8\\
    195.6 & 0.01 & & 19.1 & 1.95 & 2.34 & 32.1 & 4.9&  7.5\\
    207.1 & 0.01 & & 15.2 & 1.87 & 1.99 & 25.9 & 4.5 & 6.9\\
    214.7 & 0.04 & & 15.1 & 2.15 & 2.31 & 26.7 & 5.4 & 9.4\\
    223.9 & 0.01 & & 11.1 & 2.12 & 2.17 & 23.3 & 5.8 & 8.2\\
    229.5 & 0.14 & & 12.7 & 1.70 & 1.89 & 27.4 & 5.1 & 8.3\\
    242.2 & 0.01 & & 9.3 & 1.52 & 1.66 & 15.6 & 2.4 & 3.4\\
    345.3 & 7.48 & & 2.4 & 0.42 & 1.96 & 5.5 & 1.6 & 2.3\\
    370.3 & 7.84 & & 1.6 & 0.46 & 1.08 & 5.8 & 1.8 & 2.4\\
    464.1 & 41.9 & & 2.5 & 0.72 & 1.19 & 3.2 & 0.4 & 0.4\\
    \hline
    \end{tabular}
    \label{tab:Table_1}
\end{table*}
\section{Observations}
\label{sec:obs}

The Karl G. Jansky VLA data used in this paper were published by \citet{cho21}. 
They were obtained between days 2 and 1744 after the discovery  of nova V5668 Sgr (2015 March 15), in bands L ($1-2$ GHz), C ($4-8$ GHz), Ku ($12-18$ GHz) and Ka ($26.5-40$ GHz).  Figure \ref{fig:Fig_1} presents the light curves at the observed central frequencies (1.26, 1.74, 5.0, 7.0, 13.5, 16.5, 29.5 and 35.0 GHz). 
The systematic flux density uncertainties of the data, estimated by \citet{cho21}, were 5\% for $\nu < 10$   GHz and 10\% otherwise.
Figure \ref{fig:Fig_2}  presents the VLA spectra of V5668 Sgr on days 842 and 966 after eruption, together with the flux density at 230 GHz measured with ALMA on the intermediate day 927 \citep{dia18};   notice that, although the source was resolved with ALMA, the maximum recoverable scale angular resolution was about 4 arcsec, so that we do not expect any flux density lost in the ALMA image. Fig.  \ref{fig:Fig_2} also shows, as broken lines, the bremsstrahlung emission model, described in Section \ref{results}, fitted to the VLA data. We can see that the 230 GHz flux density lies between the two extrapolated spectra, implying that dust is not responsible for a major part of the emission at that frequency, as already pointed out by \citet{dia18} and confirmed by IR observations \citep{tak21}.

We also used the X-ray observations obtained with the X-ray Telescope (XRT) on the Neil Gehrels Swift Observatory, presented by \citet{gor21}. The data were re-analysed using the latest software and calibration files. We obtained the bolometric fluxes and temperatures corresponding to the hard, optically-thin thermal  APEC component of that X-ray emission. We converted fluxes into luminosities assuming a distance of 1.2 kpc to V5668 Sgr \citep{tak21}. The X-ray data are presented in Table \ref{tab:Table_1};  the data were binned to decrease the error bars and obtain a better value of the model temperature. The columns represent (1) day since discovery (the midpoint of the data bin), (2) bin width, (3) unabsorbed bolometric  luminosity, (4) one sigma negative uncertainty on the  luminosity (5) one sigma positive uncertainty on the  luminosity, (6), (7) and (8),  list the temperatures  and their respective negative and positive one sigma uncertainties, respectively.  Absorption was calculated assuming a  neutral hydrogen column density of $1.4 \times 10^{21}$ cm$^{-2}$  \citep{geh18}.
\section{Results}
\label{results}
\subsection{The Clump Model}
\label{Model}
Based on the high angular resolution observations at millimetre waves obtained with ALMA that revealed that the continuum emission originates in a large number of compact clumps \citep{dia18}, we modeled the observed radio spectra as the superposition of the bremsstrahlung spectra of a fixed number of spherical clumps of constant mass and composition,  distributed in a spherical shell, and determined their density and temperature by fitting the model to the observations. This model, besides being close to the observations (except for using a fixed  value for the mass of all the clumps) has the advantage of being independent of the position of the clumps in the the remnant (in all models the covering factor is less than one, the reason will be discussed in Section \ref{Parameters}). 
The bremsstrahlung spectrum of a homogeneous sphere can be calculated analytically as \citep{ost65,oln75}:
\begin{equation}
    S_\nu = \frac{2\pi kT_{\rm e}\nu^2R^2_{\rm cl}}{c^2D^2}\bigg[1-\frac{2}{\tau^2_\nu}[1-(\tau_\nu +1)e^{-\tau_\nu}]\bigg],
\end{equation}
where $T_{\rm e}$ is the electron temperature, $\nu$ the frequency, $R_{\rm cl}$ the radius
of the clump, $D$ the distance of the source and $c$ the speed of light; $\tau_\nu$ is the optical depth of each clump, which can be calculated  \citep{sut93} from:
\begin{equation}
    \tau_\nu = 0.06028n_{\rm e}^2T_{\rm e}^{-1.5}\nu^{-2}R_{\rm cl}G_{\rm ff},
\end{equation}
with
\begin{equation}
    G_{\rm ff}=\frac{n_{\rm H}}{n_{\rm e}}\sum_{\rm el,i}A_{\rm el}X_{\rm i}z_{\rm i}^2g_{\rm ff}(\nu,T_{\rm e},Z_{\rm i}).
\end{equation}
where $n_{\rm H}$ is the number density of hydrogen atoms, $n_{\rm e}$ the electron number  density, $A_{\rm el}$ the element abundance relative to hydrogen, $X_{\rm i}$ the ionization fraction of the considered ion, $z_{\rm i}$ the effective nuclear charge, including screening effects, and $g_{\rm ff}$ the Gaunt factor.
Accurate values of the Gaunt factor and their dependence on
frequency and temperature are required to obtain reliable spectra
covering both cm and mm wavelengths. These values were obtained
by interpolation of the recently published tables, calculated with
high degree of accuracy by \citet{hof14}, resulting in the following
expression, valid for $(0 < \log \gamma^2 < 2)$ and $(-6 < \log u < -4)$:
\begin{equation}
    g_{\rm ff}(\nu,T_{\rm e},Z)=m(\gamma^2)\log u + b(\gamma^2),
\end{equation}
with
\begin{equation}
    m(\gamma^2)=-1.265+0.010\log \gamma^2,
\end{equation}
and 
\begin{equation}
    b(\gamma^2)=-0.0539-0.5467\log \gamma^2,
\end{equation}
where
\begin{equation}
    u=\frac{h\nu}{kT_{\rm e}},
\end{equation}
and
\begin{equation}
    \gamma^2=\frac{Z^2{\rm Ry}}{kT_{\rm e}},
\end{equation}
$\rm Ry$ is the Rydberg constant.
\subsection{Exploring the parameter space}
\label{Parameters}

Although the number of free parameters in the model is large and time dependent, some of them were already obtained by \citet{tak21}, modeling optical and infrared spectra at epochs close to the VLA and ALMA observations. From their data, we fixed the electron temperature as $1.21 \times 10^4$ K, hydrogen and helium mass abundances as $X = 0.45$ and $Y = 1 -X$, respectively; we also
consider the hydrogen and helium atoms totally ionized, 73\% of
helium in the form of He$^+$ and the rest He$^{++}$.  The assumption $Z=0$ was only used for the calculation of the radio spectra of the clumps, since the contribution of the heavy elements to bremsstrahlung emission is  small at temperatures of $10^4$ K, but they  were taken into account for the calculation of the X-ray luminosity, as discussed in Section \ref{X-rays}.

To explore the parameter space, we fixed the abundances and total mass $M_{\rm tot}$ of the clumps:
\begin{equation}
    M_{\rm tot}=N_{\rm cl}\frac{4}{3}\pi R_{\rm cl}^3 n_{\rm H}m_{\rm H} \bigg(1+\frac{Y}{X}\bigg) , 
\end{equation}
and determined the value of the radius and electron temperature that
best fitted the spectrum of day 966, for several values of the number
of clumps $N_{\rm cl}$ (20, 30, 40, 50, 70 and 100);  the density of the clumps was determined from their radius and  mass.  There is a unique set of model parameters that fit the data, because the optically thick part of the spectrum depends on temperature and radius, while the optically thin part depends on temperature, radius and density.
We also used the expansion velocity of the shell given by \citet{tak21},  $v_{\rm sh}= 590$ km s$^{-1}$ to calculate the covering factor $f$:
\begin{equation}
    f=N_{\rm cl}\bigg(\frac{R_{\rm cl}}{R_{\rm sh}}\bigg)^2,\
\end{equation}
where $R_{\rm sh}$ is the radius of the shell, given by:
\begin{equation}
    R_{\rm sh}=v_{\rm sh} t.
\end{equation}
 From the best fitting models we obtained relationships between $T_{\rm e}$, $M_{\rm tot}$, $R_{\rm cl}$ and $f$ for each value of $N_{\rm cl}$. We found power law relations between them, for each value of $N_{\rm cl}$. Graphs of these relations are presented in  Appendix \ref{Ap_A}. The results can be summarized in the following expressions: 
\begin{equation}
    \log p_{\rm i}=M_{\rm ij}(N_{\rm cl})\log p_{\rm j}+N_{\rm ij}(N_{\rm cl}),
\end{equation}
with
\begin{equation}
    M_{\rm ij}(N_{\rm cl})=A_{\rm ij}\log N_{\rm cl}+B_{\rm ij},
    \label{Mij}
\end{equation}
and
\begin{equation}
    N_{\rm ij}(N_{\rm cl})=C_{\rm ij}\log N_{\rm cl}+D_{\rm ij},
    \label{Nij}
\end{equation}
where $p_{\rm i}$ represents the parameters $T_{\rm e}$, $R_{\rm cl}$, and $f$ , and $p_{\rm j}$ the parameters $M_{\rm tot}$ and $T_{\rm e}$. The values of the coefficients $A_{\rm ij}$, $B_{\rm ij}$, $C_{\rm ij}$
and $D_{\rm ij}$ are presented in Table \ref{tab:Table_2} .
\begin{figure}
\includegraphics[width=\columnwidth]{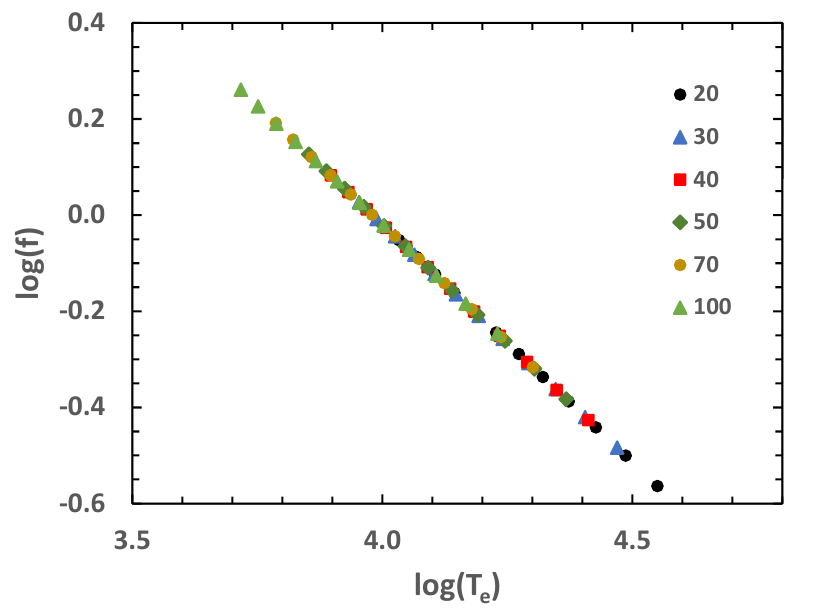}
	\caption{Covering factor as a function of the electron temperature  for different values of the total number of clumps, shown as different symbols, defined at the right side of the graph. }
    \label{fig:Fig_3}
\end{figure}
\begin{figure}
\includegraphics[width=\columnwidth]{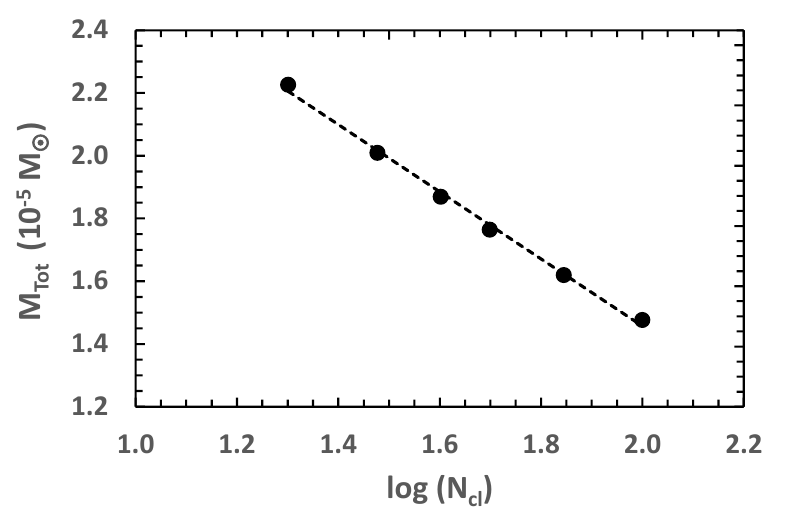}
	\caption{Relationship between the total mass of the ejecta and the number of clumps, for the models with electron temperature $1.21 \times 10^4$ K that best fit the spectrum of day 966. }
    \label{fig:Fig_4}
\end{figure}
\begin{figure*}
\includegraphics[width=17cm]{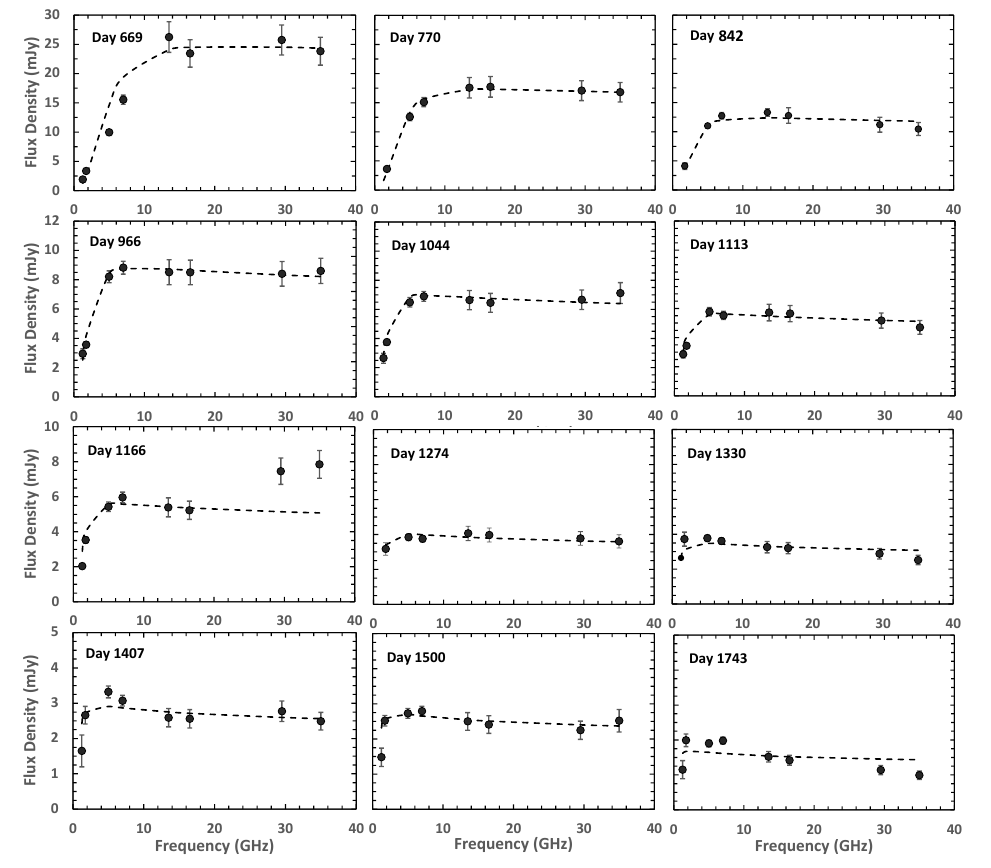}
	\caption{Spectra of V5668 Sgr, obtained with the VLA (black dots) and best fit of  models consisting of 40 clumps with constant temperature of $1.21\times 10^4$ K   (dashed line) }
    \label{fig:Fig_5}
\end{figure*}

\begin{figure*}
\includegraphics[width=17cm]{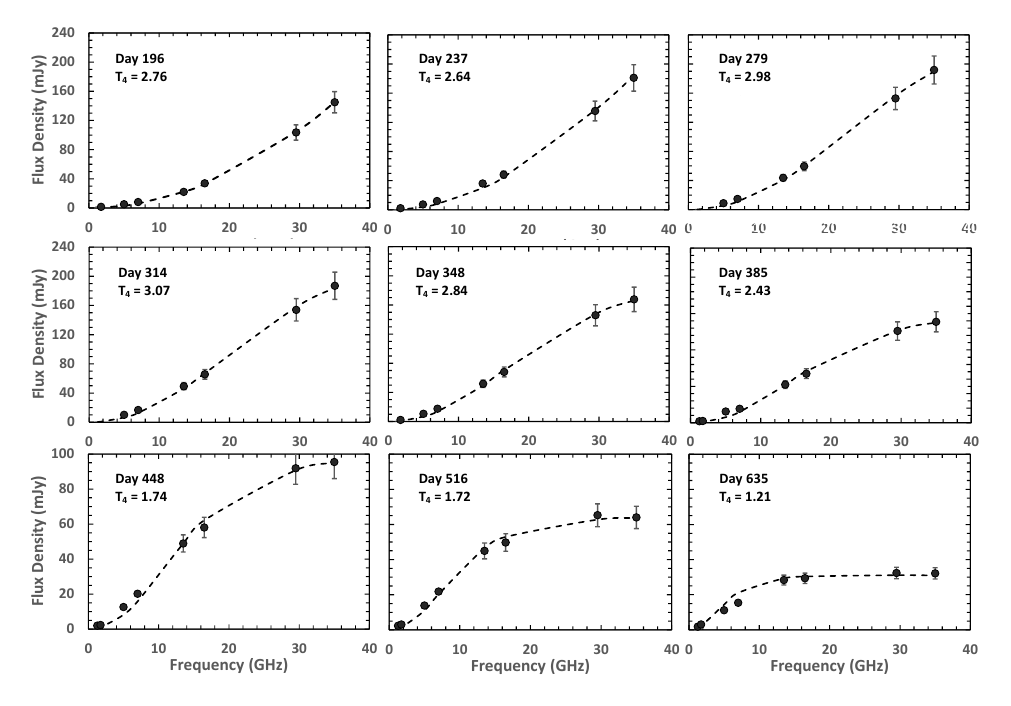}
	\caption{Spectra of V5668 Sgr, obtained with the VLA (black dots) and best fit of  models with  40 clumps and variable temperature (dashed line). The temperature of each model, in units of $10^4$ K, is shown in the upper left corner. }
    \label{fig:Fig_6}
\end{figure*}


\begin{table}
    \centering
     \caption{Coefficients of the relationships given in equations \ref{Mij} and \ref{Nij} }
    \begin{tabular}{l|c|c|c|c|c}
    \hline
 $p_{\rm i}$ & $p_{\rm j}$ & $A_{\rm ij}$ & $B_{\rm ij}$ & $C_{\rm ij}$ & $D_{\rm ij}$\\
 \hline
 $T_{\rm e}$  &  $M_{\rm tot}$ & -1.75e-2 &  1.851 & -3.82e-1 & -3.927\\
 $R_{\rm cl}$ &  $T_{\rm e}$   &  7.94e-4 & -0.496 & -5.03e-1 & 17.648\\
 $f$          &  $T_{\rm e}$   &  1.59e-3 & -0.992 & -6.72e-3 &  3.949\\
    \hline
    \end{tabular}
    \label{tab:Table_2}
\end{table}  

\begin{figure*}
\includegraphics[width=15cm]{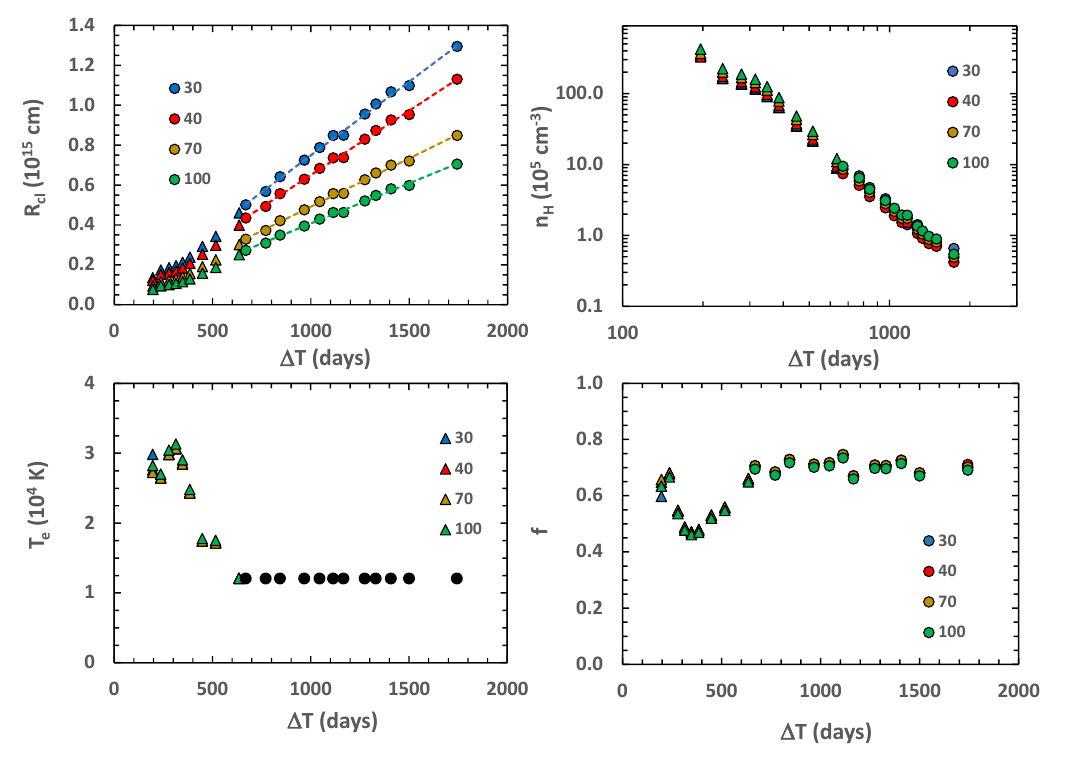}
	\caption{Radius, hydrogen number density, electron temperature and covering factor of the models that best fitted the spectra as a function of time elapsed since eruption of nova V5668 Sgr (top left, top right, bottom left and bottom right, respectively) for different values of the number of clumps (30, 40, 70 and 100)}
    \label{fig:Fig_7}
\end{figure*}

\begin{figure*}
\includegraphics[width=15cm]{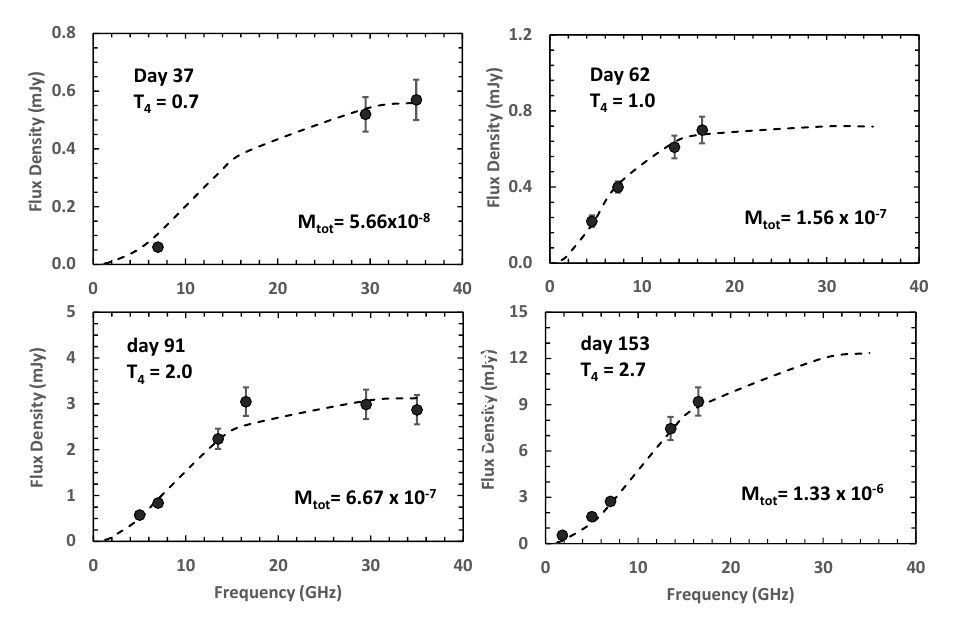}
	\caption{Spectra of V5668 Sgr, obtained with the VLA (black dots) and best fit of  models with  40 clumps and variable temperature and mass (dashed line). The temperature of each model, in units of $10^4$ K, is shown in the upper left corner, and the total mass of the ejecta in the lower right corner. }
    \label{fig:Fig_6b}
\end{figure*}


\begin{figure}
\includegraphics[width=9cm]{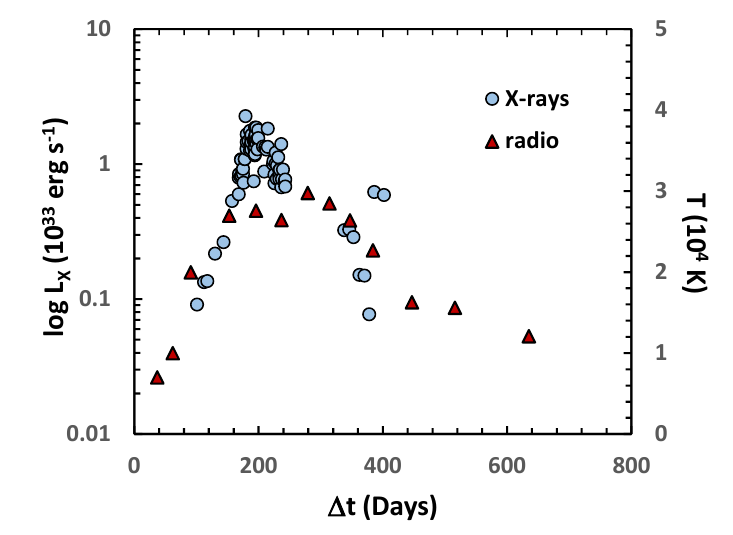}
	\caption{Observed super-soft X-ray luminosity,  (blue circles, left axis) and  temperature of the clumps, obtained from the models (red triangles, right axis) as a function of time}
    \label{fig:Fig_7b}
\end{figure}

\begin{figure*}
\includegraphics[width=17cm]{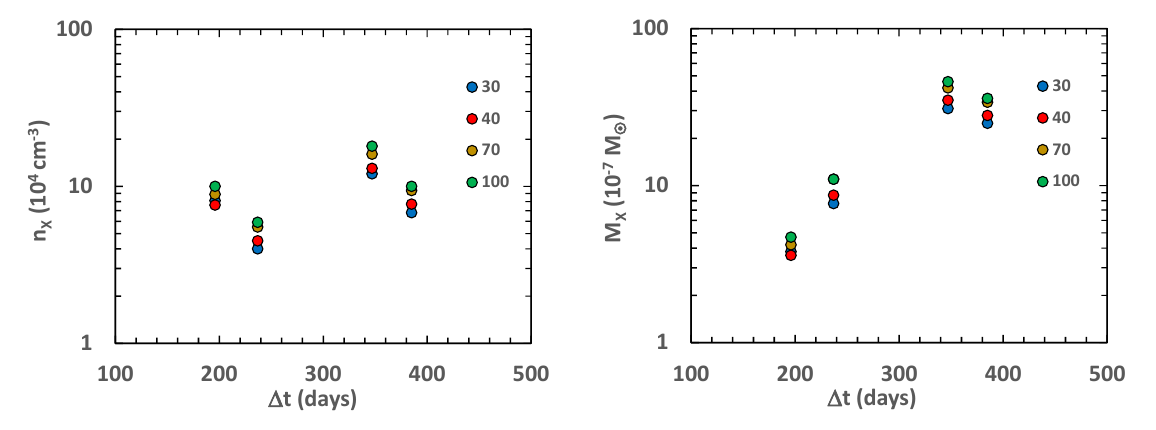}
	\caption{Left panel:number density of the hot plasma component of V5668 Sgr, at the epochs in which almost simultaneous X-ray and radio data are available. Right panel: total mass of the hot gas in the shell for the same epochs.}
    \label{fig:Fig_8}
\end{figure*}

\begin{figure*}
\includegraphics[width=17cm]{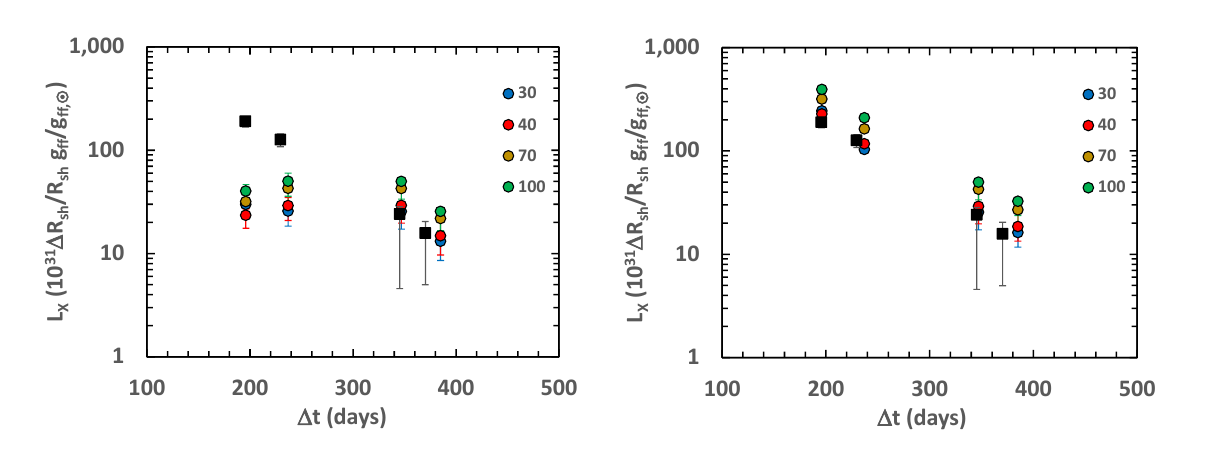}
	\caption{Observed (black squares) and model (1-10) keV luminosity (circles) of V5668 Sgr for the epochs at which almost simultaneous X-ray and radio data are available. Left panel: models use the masses presented in the right panel of Figure \ref{fig:Fig_8}. Right panel: all models are calculated with the highest mass, corresponding to day 345}
    \label{fig:Fig_9}
\end{figure*}
\begin{figure}
\includegraphics[width=8cm]{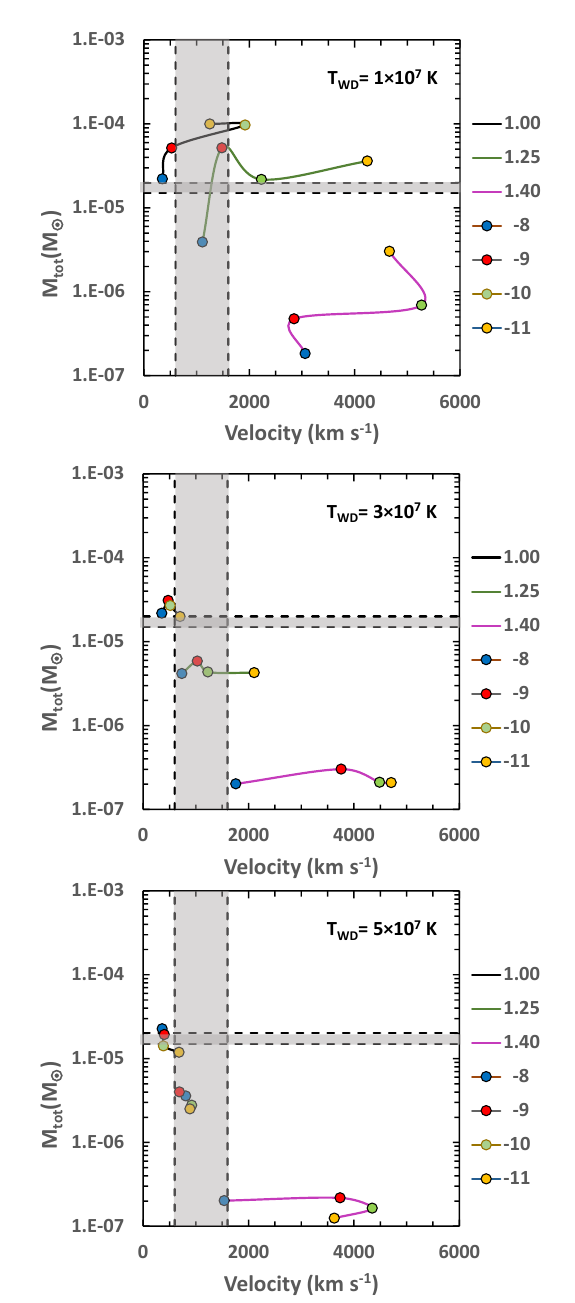}
	\caption{Top panel: maximum velocity of a nova ejecta as a function of the ejected mass, for three values of the white dwarf mass (continuous line) of central temperature $10^7$ K, and four values of the accretion rate (coloured circles). Middle and bottom panels: the same as the top panel, for central temperatures 3 and $5 \times 10^7$ K. Gray area represent the observed values of the velocity of the nova ejecta and total mass of the clumps obtained from the  models.}
    \label{fig:Fig_10}
\end{figure}
 An interesting result of these relationships is that the dependence of the covering factor on the electron temperature is independent of the number of clumps considered in the model,  as can be seen in Figure \ref{fig:Fig_3}. This is easy to understand, because the number of clumps is given by the ratio between the total flux density of the shell and that of the individual clump, so that:
\begin{equation}
    f(N_{\rm cl}) =N_{\rm cl}\bigg(\frac{R_{\rm cl}}{R_{\rm sh}}\bigg)^2=\frac{S_{\rm tot}}{S_{\rm cl}}\bigg(\frac{R_{\rm cl}}{R_{\rm sh}}\bigg)^2\propto S_{\rm tot} T_{\rm e}^{-1},
\end{equation}

\noindent
 since $S_{\rm cl} \propto T_{\rm e}R_{\rm cl}^2$ in the optically thick part of the spectrum.

For the selected value of the electron temperature, $T_{\rm e} = 1.21 \times 10^4$ K, the relationship between the total mass of the ejecta $M_{\rm tot}$ and the number of clumps is presented in Figure 4, it can be expressed analytically as:
\begin{equation}
    M_{\rm tot}(N_{\rm cl},T_{\rm e}=1.21\times 10^4 {\rm ~K})=(3.6-1.07\log N_{\rm cl})\times 10^{-5} {\rm ~M_{\odot}}
    \label{mass}
\end{equation}
The total mass varied between $2.0\times 10^{-5}$ and $1.5\times 10^{-5}$ M$_\odot$ for $N_{\rm cl}$ ranging from 30 to 100 (the number of clumps is arbitrary, but compatible with the number detected with ALMA); in all cases, the total mass is smaller than that obtained by \citet{tak21}, the implication
of this result is discussed in Section \ref{WD mass}. The radius of the clumps varied between $7.2\times 10^{14}$ and $3.9\times 10^{14}$ cm and the hydrogen density between $2.3\times 10^5$ and $3.1\times 10^5$ cm$^{-3}$. These values are compatible with those observed in the clumps detected with ALMA on day 927 \citep{dia18}.

\subsection{Time evolution of the clumps}
\label{Evolution}
 From the results obtained in Section \ref{Parameters} for day 966, we fixed the parameter $T_{\rm e}=1.21 \times 10^4$ K and $M_{\rm tot}=M_{\rm tot}(N_{\rm cl},T_{\rm e})$, given by equation \ref{mass},  which determines the value of the mass of each clump $M_{\rm cl}=M_{\rm tot}/N_{\rm cl}$. We obtained    the radius of the clumps for all the other epochs, by fitting the model to the observed spectrum for different values of $N_{\rm cl}$ (30, 40, 70, 100); notice that the density of the clumps is determined from their mass and radius. We were able to find satisfactory fittings for all epochs after $ t = 635$  days. The observed spectra and the fitting models for $N_{\rm cl}=40$ are presented in Figure \ref{fig:Fig_5}. We can see that the agreement is very good for all days except for day 1166 in which we did not use the two higher frequencies for the fitting.  We need to emphasize the importance of this result, because the radius of the clumps are completely determined by the optically thick part of the spectrum,  so that the fact that the model fits also the optically thin part of the spectrum, which depends on radius and  density, means that the assumption that all the clumps have the same mass and temperature at all these epochs is valid.

For epochs previous to $t = 635$  days, it was necessary to vary also the electron temperature to fit the data, the results for $N_{\rm cl}=40$ are presented in Figure \ref{fig:Fig_6}, where the values of $T_{\rm e}$ are also shown together with the spectra.

In Figure \ref{fig:Fig_7} we present the values of the physical parameters of the clumps ($R_{\rm cl}, n_{\rm H}, T_{\rm e}$ and $f$) as a function of the epoch of observation, for different values of $N_{\rm cl}$. We can see that $R_{\rm cl}$ increased linearly with time for the epochs in which the temperature was constant, with an expansion velocity that depends on $N_{\rm cl}$. The temperature of the clumps at the early times of the nova evolution are almost independent of the number of clumps considered, and the covering factor varied with the temperature, as it was explained in Section 4.2;  it was always smaller than 0.7, implying that it was not necessary to consider additional absorption by superposition of clumps along the line of sight.  The volume filling factors $f_{\rm V}$ can be obtained from $f_{\rm V}=fR_{\rm cl}/R_{\rm sh}$; their maximum values varied from 0.11 to 0.06 for $N_{\rm cl}$ varying from 30 to 100, in agreement with the  estimated value $f_{\rm V}<0.2$ obtained from ALMA data \citep{dia18}.  The density of the
clumps varied as a power law of the elapsed time, $n_{\rm H}\propto  t^{-\alpha}$, with $\alpha = 2.985 \pm 0.004$.
This result will be discussed in Section \ref{WD mass}.
\section{Discussion}
\label{Discussion}

In Section \ref{results} we showed that the radio spectra obtained with the VLA can be explained, even at the early times of the nova evolution, by the superposition of the bremsstrahlung spectra of a large number of small  $(<\sim 10^{15}$ cm) and dense ($\sim 10^7$ cm$^{-3}$) clumps.  The existence of a large number of clumps  is in agreement with the scenario in which the clumps are formed in radiative shocks, originated by the interaction between the accreted material ejected during the
nova eruption, and the faster radiation-driven wind of the white dwarf that emerged later. In other novae, these shocks produced initially high energy $\gamma$-rays, coincident with optical peaks \citep{ayd20a}.  That was not evident in V5668 Sgr because the $\gamma$-ray emission was very weak, many times below the detection limit \citep{gor21}. The origin of the shocks in classical novae is different from those in recurrent novae, in which they originate from the collision of the ejected matter and the wind of the red giant companion \citep{kan07,kan16}.
The radio emission  V5668 Sgr, was first detected at only one frequency on day 23 (0.12 mJy at 29.5 GHz), and at three frequencies on day 37; on each occasion, it was also detected at $\gamma$-rays \citep{che16}. The last high energy detection occurred on day 62, although the observations lasted for 77 days.

The VLA spectra, obtained on days  37, 62, 91 and 153  could be explained by bremsstrahlung emission of the same number of clumps as those used to fit later spectra, but with smaller ionized mass, which makes sense, since at that time,  the white dwarf was still contributing with its wind to the ejected mass, and also  part of the gas was condensed into dust grains that blocked part of the ionizing photons, as shown by the dip in the $B$, $V$, $R$ and $I$  light curves and the increase in the near infrared colors \citep{geh18}; the observed spectra and the best fitting models are presented in Fig. \ref{fig:Fig_6b}. 
 Notice that none of the spectra present a spectral index at the lower frequencies compatible with non-thermal emission.
The minima in the optical light curves was close to day 110 after eruption and the end of the dust destruction, close to day 196, after which the mass of our models remained constant. 
For $N_{\rm cl}= 40$, the masses obtained for days  37, 62, 91
and 153 were $5.7\times 10^{-8}$, $1.6\times 10^{-7}$, $6.7\times 10^{-7}$ and $1.3\times 10^{-6}$ M$_\odot$, with electron temperatures of $7.0 \times 10^3$, $1.0 \times 10^4$, $2.0 \times 10^4$ and $2.7 \times 10^4$ K, respectively. The  total mass of ionized gas for the later observations was $1.9\times 10^{-5}$ M$_\odot$. The fitted spectra can be seen in Fig. \ref{fig:Fig_6b}.

 The temperature of the clumps increased from day 37 to 237, when it reached $3.1 \times 10^4$ K and started a systematic decrease,   until it attained  the equilibrium temperature of $1.21\times 10^4$ K on day 635, as can be seen in Fig. \ref{fig:Fig_7}.  This variation follows exactly the behavior of the super soft X-ray luminosity \citep{geh18,gor21}, and confirms the assumption of photoionization as the heating source of the nova ejecta  \citet{cun15}.  Fig. \ref{fig:Fig_7b} presents the observed luminosity of the super-soft X-ray component  and the clump´s temperatures as a function of time.
\subsection{ The two component model and X-ray luminosity}
\label{X-rays}

After day $t = 635$, all the spectra of V5668 Sgr could be reproduced by models with constant mass and electron temperature.  The radius of the
clumps increased at a rate that varied from  85 to 46 km s$^{-1}$ for $N_{\rm cl}$ between 30 and 100, and their density decreased as $n_{\rm H}\propto t^{-3.0}$ for all values of $N_{\rm cl}$. The exponent  3.0 in the power law implies that the decrease in density is proportional to the increase of the volume $V_{\rm sh}$    in which the clumps are distributed, which we assume it is a shell of relative width $(\Delta R_{\rm sh}/ R_{\rm sh})$:
\begin{equation}
\label{volume}
    V_{\rm sh} \sim  4\pi R_{\rm sh}^3(\Delta R_{\rm sh}/ R_{\rm sh}),
\end{equation}
We assumed that the expansion velocity  and the relative size of the shell  remain constant, since $R_{\rm sh} = v_{\rm sh} t$.  In fact, the spectral data of V5668 Sgr in \citet{tak21} indicate that the expansion velocity of the nova is constant at least from ~500 days to ~900 days after eruption. Earlier than that, \citet{ban16} have a similar velocity estimate for day ~65 after eruption, suggesting that the expansion velocity does not vary significantly in the period of our models.  During the early evolution of the shell, large velocity variations are unlikely to occur given the relatively small interaction with the Interstellar Medium. 

Based on these results, we suggest that the clumps are in pressure equilibrium with a hotter and thinner  medium, in which they are immersed. The hot  medium represents the expanding material of the shock that was not condensed into clumps.
A similar  two component model was
proposed by \citet{sai74}  to explain the blue featureless continuum in a declining classical nova, although the origin of the two media were not discussed because $\gamma$-rays, emitted by the shocked material, were only discovered later.

Large density gradients are required in order to explain the ionization range observed in the emission line spectrum of novae. The role of condensations in the photoionization modeling of nova shells was first explored by \citet{wil92}. Using a density dependent filling factor with a Kolmogorov-type distribution, he found that the shell quickly evolves to matter bounded, with  regions of low ionization confining  globules cores. The density range for ionized condensations in these filling factor models is consistent with the derived  densities and temperatures for the clumps of V5668 Sgr  ($\sim 2 \times 10^{4} - 3 \times 10^{7}$) cm$^{-3}$ and $(1.2 - 3.1) \times 10^4$ K, as seen in Figure 7.

The model proposed here allows us to calculate the density of the hot component for each epoch, if its electron temperature is known, since pressure equilibrium implies:
\begin{equation}
\label{density}
    n_{\rm X}T_{\rm e,X}=n_{\rm cl}T_{\rm e,cl},
\end{equation}
where $n_{\rm cl}$ and $n_{\rm X}$ are the number densities of the clumps and of the
hot material and, $T_{\rm e,cl}$ and $T_{\rm e,X}$ their respective electron temperatures.

 The hot material is responsible for the hard X-ray luminosity, which can be calculated from:
\begin{equation}
\label{L_X}
    L_{\rm X}=5.86\times 10^{-27}n_{\rm ion,X}n_{\rm e,X}T_{\rm e,X}^{1/2}{\rm e}^{-1/T_{\rm e,X}{\rm (keV)}}R_{\rm sh}^3\frac{\Delta R_{\rm sh}}{R_{\rm sh}}\frac{g_{\rm ff+fb}}{g_{\rm ff+fb,\odot}},
\end{equation} 
where $n_{\rm ion,X}$ and $n_{\rm e,X}$ are, respectively, the number density of ions and electrons of the hot gas,   and $n_{\rm ion,X}+ n_{\rm e,X} = n_{\rm X}$; $g_{\rm ff+fb}$ is the Gaunt factor, which includes both free-free and free bound transitions, the last one depends strongly on the plasma composition, specially the C and O abundances. $g_{\rm ff+fb,\odot}$ is the corresponding gaunt factor for the solar composition, which for temperature of $5\times 10^6$ K has the value 7.9 \citep{sut93}.

 To obtain the expected X-ray luminosities  of the hot shell at different epochs and compare them with the observations,  it is necessary to know the electron densities and temperatures. The temperatures were obtained from the APEC spectral fitting,  presented in Table \ref{tab:Table_1}, and the number densities were  calculated from equation \ref{density} at the four epochs at which both radio and the X-ray spectra were obtained almost simultaneously: $ t= 196, 230, 345$ and 370.
 Figure \ref{fig:Fig_8}  shows the  calculated   densities of the hot material (left panel), and the total mass of the shell (right panel), assuming the shell volume given by equation \ref{volume} with $\Delta R_{\rm sh}/R_{\rm sh} =1$, for four number of clumps  (which define the  values of $n_{\rm X}T_{\rm e,X}$).   We can see that the masses are much smaller than the total mass of the clumps, and that for the  first two epochs, they are one order of magnitude smaller than those obtained for the two last ones.

The left panel of Figure \ref{fig:Fig_9} presents the values of $L_{\rm X}$  obtained from the observations and from our model, in units of $10^{31}(\Delta R_{\rm sh}/R_{\rm sh})(g_{\rm ff+fb}/g_{\rm ff+fb,\odot}$)  erg s$^{-1}$.  We can see that the agreement between them is good at the two last epochs for all  number of clumps, but is one order of magnitude smaller than the observations for the firs two epochs.

 To see if the difference  in X-ray luminosity between model and observations  at the first two epochs is related to the fact that the masses obtained for the shells were smaller than those of the last two epochs,  we recalculated the X-ray luminosity of our models assuming the same mass at all epochs (the highest mass, corresponding to day 345).  Since the densities are fixed by equation \ref{density}, the radius of the shell was increased to obtain the right mass. The results are presented in the right panel of Figure \ref{fig:Fig_9}, where we can see an excellent agreement between models and observations, from which we conclude that the mass of the hot component was constant during the period investigated.
The larger radius necessary to obtain the model mass for days 196 and 230 implies an expansion velocities  of 1270 and 946 km s$^{-1}$, respectively. These velocities are within the range of velocities measured at early times by \citet{taj16},  closer to the velocity of the wind of the white dwarf, which is decelerated after passing through the shock.
\subsection{A distribution of clump sizes}

In the previous sections we showed that it is possible to reproduce the radio spectra of  nova V5668 Sgr  as the sum of the spectra of a number of identical spherical clumps of fixed  mass, and found a relationship between the radius and density of the clumps by fitting the model spectra to the observations  for each epoch. We also found that the variation of the density of the clumps as they expand suggests the existence of a hotter and thinner medium in which the clumps are embedded.
However, ALMA data \citep{dia18} showed that not all the clumps have the same size, and that most of them are unresolved. Therefore, we investigated the effect of having clumps of different sizes on the resulting model spectra, assuming a power law distribution of sizes of the form:
\begin{equation}
\frac{dN(R_{\rm cl})}{dR_{\rm cl}} \propto R_{\rm cl}^{\alpha},
\label{eqdNdR}
\end{equation}
\noindent
where $N(R_{\rm cl})$ is the number of clumps with radius between $R_{\rm cl}$ and $R_{\rm cl}+dR_{\rm cl}$, $R_{\rm cl}^{\rm min} \leq R_{\rm cl} \leq R_{\rm cl}^{\rm max}$ and $\alpha <0$.
We can write equation \ref{eqdNdR} as:
\begin{equation}
 \frac{dN(R_{\rm cl})}{dR_{\rm cl}}= 
\bigg( \frac{dN(R_{\rm cl})}{dR_{\rm cl}}\bigg)_{R_{\rm cl} = R_{\rm cl}^{\rm max}}\biggl(\frac{R_{\rm cl}}{R_{\rm cl}^{\rm max}}\biggl)^{\alpha}.
 \label{eq_a}
\end{equation}
For the frequencies  in which the spectrum is optically thick, the flux density of each clump $S_{\rm cl}(\nu_{\rm thick},R_{\rm cl})$ is independent  of the density and can be written as:
\begin{equation}
    S_{\rm cl}(\nu_{\rm thick}, R_{\rm cl})=S(\nu_{\rm thick}, R_{\rm cl}^{\rm max})\biggl(\frac{R_{\rm cl}}{R_{\rm cl}^{\rm max}}\biggl)^2.
    \label{eq_b}
\end{equation}

The total flux density of the model $S_{\rm tot}(\nu_{\rm thick})$ is:
\begin{equation}
        S_{\rm tot}(\nu_{\rm thick})=\int_{R_{\rm min}}^{R_{\rm max}} \frac{dN(R_{\rm cl})}{dR_{\rm cl}} S_{\rm cl}(\nu_{\rm thick}, R_{\rm cl}) dR_{\rm cl}. 
\label{eq_c}
\end{equation}
Substituting  eqs. \ref{eq_a} and \ref{eq_b} in eq. \ref{eq_c} and assuming  $R_{\rm cl}^{\rm min} << R_{\rm cl}^{\rm max}$ we obtain:
\begin{equation}
 S_{\rm tot}(\nu_{\rm thick})= \frac {R_{\rm cl}^{\rm max}}{3+\alpha} \bigg( \frac{dN(R_{\rm cl})}{dR_{\rm cl}}\bigg)_{R_{\rm cl} = R_{\rm cl}^{\rm max}} S_{\rm cl}(\nu_{\rm thick}, R_{\rm cl}^{\rm max}). 
 \label{eq_d}
\end{equation}
From the previous models in which we assumed a fixed number $N^*$ of identical clumps of radius $R^*_{\rm cl}$, we can write:
\begin{equation}
  S_{\rm tot}(\nu_{\rm thick})=N^*  S^*_{\rm cl}(\nu_{\rm thick}, R^*_{\rm cl})
  \label{eq_e}
\end{equation}
 
    Let us choose the value of $N^*$ that satisfies $R^*_{\rm cl}=R_{\rm cl}^{\rm max}$; since the spectra depend only on the radius,  $S^*_{\rm cl}(\nu_{\rm thick}, R^*_{\rm cl})=S_{\rm cl}(\nu_{\rm thick}, R_{\rm cl}^{\rm max})$.
From eqs. \ref{eq_d} and \ref{eq_e} we obtain:

\begin{equation}
   \bigg( \frac{dN(R_{\rm cl})}{dR_{\rm cl}}\bigg)_{R_{\rm cl} = R_{\rm cl}^{\rm max}}=\frac{(3+\alpha)N^*}{R^*_{\rm cl}}
   \label{h}
\end{equation}

Let us consider now the optically thin part of the spectrum, which depends on both density and radius. The density $n_{\rm H }$ must be the same for all the clumps because they are in equilibrium with the external medium. Therefore:

\begin{equation}
    S_{\rm cl}(\nu_{\rm thin}, n_{\rm H}, R_{\rm cl})=S_{\rm cl}(\nu_{\rm thin}, n_{\rm H}, R_{\rm cl}^{\rm max})\biggl(\frac{R_{\rm cl}}{R_{\rm cl}^{\rm max}}\biggl)^3.
    \label{eq_f}
\end{equation} 

The total integrated flux density will br:

\begin{equation}
 S_{\rm tot}(\nu_{\rm thin})= \frac {R_{\rm cl}^{\rm max}}{4+\alpha} \bigg( \frac{dN(R_{\rm cl})}{dR_{\rm cl}}\bigg)_{R_{\rm cl} = R_{\rm cl}^{\rm max}} S_{\rm cl}(\nu_{\rm thin}, R_{\rm cl}^{\rm max}). 
 \label{eq_g}
\end{equation}
Using  again the model of identical clumps:
\begin{equation}
  S_{\rm tot}(\nu_{\rm thin})=N^*S_{\rm cl}(\nu_{\rm thin}, n^*_{\rm H}, R^*_{\rm cl})
  \label{eq_h}
\end{equation}
From eqs.  \ref{eq_e},  \ref{eq_f} and \ref{eq_h}, we obtain:
\begin{equation}
        S_{\rm cl}(\nu_{\rm thin}, n_{\rm H}, R_{\rm cl}^{\rm max}) = \frac{4+\alpha}{3+\alpha}S_{\rm cl}(\nu_{\rm thin}, n^*_{\rm H}, R^*_{\rm cl}),
\end{equation}
\noindent
or
\begin{equation}
    n_{\rm H}=\bigg(\frac{4+\alpha}{3+\alpha}\bigg)^{1/2} n^*_{\rm H}
\label{eq_j}
\end{equation}
Therefore, the model in which the distribution of clump sizes is a power law, is equivalent to the model in which all clumps are identical, when their radii are equal to the maximum observed radius and the densities  related are by equation \ref{eq_j}.

In the case of the ALMA observations, the maximum radius was of the order of $7 \times 10^{14}$ cm \citep{dia18}, similar to the radius of the $N^*=40$ identical clumps.

\subsection{Mass of the white dwarf}
\label{WD mass}
Numerical models of thermonuclear explosions in accreting white dwarfs showed that the physical and dynamical properties of the evolving novae depend on the white dwarf mass $M_{\rm WD}$, temperature $T_{\rm WD}$ of its isothermal core, and accretion rate from the companion star $\dot{M}_{\rm acc}$, \citep{pri95},  although $T_{\rm WD}$ could also be related to $\dot{M}_{\rm acc}$ \citep{tow04}.
A large number of models are available, covering the complete range of the three dimensional parameter space ($M_{\rm WD}$, $T_{\rm WD}$, $\dot{M}_{\rm acc}$) that led to the observed properties of nova remnants: their total ejected mass $M_{\rm tot}$, maximum  velocity  of the ejecta $v_{\rm max}$ and   brightness decrease rate after the maximum in the optical light curve \citep{yar05}.

The total mass of the ejecta obtained by fitting radio light curves to the different models described in Section \ref{Introduction} turned out to be larger than that predicted by the accreting white dwarf models,  when related to the brightness decrease rate \citep{roy12}.

 In the case of nova V5668 Sgr, we will compare the relationship between the total ejected mass $M_{\rm tot}$ obtained from our models and the maximum observed velocity of the ejecta $v_{\rm max}$, because the optical light curve was contaminated by the formation of dust and possibly also by the contribution of optical emission from the shocks. 
In Figure \ref{fig:Fig_10} we present the relation between these two parameters for three values of the white dwarf core temperature (1, 3 and 5 $ \times 10^7$ K), three values of the white dwarf mass (1, 1.25 and 1.4  M$_\odot$) and four values of the accretion rate ($10^{-8}, 10^{-9}, 10^{-10}$, and $10^{-11}$  M$_\odot$  y$^{-1}$), obtained from the accreting white dwarf models \citep{yar05}, together with the total mass of our models,  and assuming as the possible values of the maximum velocity of the ejecta those measured by \citet{taj16}, which include the velocity used in our model. 
By comparing the two sets of models,  we can see that they agree for a white dwarf mass of about 1.25M$_\odot$ with a central temperature of $10^7$ K and accretion rate between $10^{-9}$ and $10^{-8}$  M$_\odot$  y$^{-1}$.

\section{Conclusions}
\label{Conclusions}
In this work we analysed the origin and evolution of the radio spectra on nova V4668 Sgr obtained with the VLA between frequencies 1.2 and 35 GHz \citep{cho21}.

Based on the high resolution observations of  V5668 Sgr obtained with ALMA 927 days after eruption \citep{dia18}, which showed that the expanding shell of material ejected by the nova eruption is formed by compact clumps, we propose that these  clumps were formed at early times in the nova evolution. The probable origin of these clumps would be radiative instabilities in the shocks formed by collisions of fast winds and the slower expanding shell.

To verify this assumption, we first fitted the observed VLA spectrum of day 966, close to the ALMA observation, with the superposition of the spectra of a fixed number (20, 30, 40, 50, 70, 100) of spherical homogeneous clumps of the same physical properties (mass, temperature, radius). For each number of clumps we found  relationships between the different physical  parameters of the clumps that fitted the observations. 
We fixed their temperature to the value obtained by \citet{tak21} and found a relationship between the total mass of the ejecta and the number of clumps.
Using these masses and the fixed temperature, we verified that the model was able to reproduce satisfactorily all  the radio spectra observed after day 635. To fit earlier data, it was necessary to use different temperatures for each epoch  
between days 196 and 635, and also different total masses, between days 37 and 153.   However, no indication of non-thermal emission was found from the spectral indices of the low frequency emission.
We found that after day 635, the density of the clumps decreased as the third power of the elapsed time after eruption, which means, assuming constant expansion velocity, as the increase of the volume of the shell (assuming a constant ratio between its width and radius).

We  suggest that the clumps are immersed, and in pressure equilibrium, in a hotter and thinner expanding   medium.

We calculated the density of this medium at four epochs in which X rays were detected, using the temperatures inferred  from the hard X-ray spectra and found that it was possible to reproduce the intensity of the X-ray emission with this material at  two of the epochs: days 345 and 370 after eruption. For the two earlier epochs: days 196 and 230, it was necessary to assume a larger emitting volume to fit the  X-ray luminosity, which required a larger expansion velocity: 1270  and 946 km s$^{-1}$, respectively, which are also compatible with observations \citep{taj16}.  Finally, we showed that the model also applies to a power law distribution of clump sizes, when the radius of the identical clumps is equal the the maximum observed radius.

 Hydrodynamic models predict that clumps can be created by different processes in the early phases of a classical nova eruption  \citep{cas11,ste19} but we often lack supporting observational data to pinpoint the moment of the clumps' formation. As it is not possible to resolve small structures until they have expanded, we can only infer the clumps' presence indirectly close to the outburst. Our models provide evidence that the clumps are formed as early as (if not earlier than) 196 days after eruption in the case of V5668 Sgr, and that they survive at least until 1743 days after eruption.

The total mass of the ejecta turned out to be smaller than that predicted by \citet{tak21} assuming an homogeneous shell with a radial density distribution, and agree with evolutionary models of novae, with mass of 1.25 M$_\odot$, central temperature of 10$^7$ K and accreting mass from the companion star at a rate of of $10^{-9} - 10^{-8}$ M$_\odot$ y$^{-1}$.

\section*{Acknowledgements}
 
ZA acknowledges Brazilian agencies FAPESP (grant \#2014/07460-0) and CNPq (grant \#304242/2019-5). MD thanks the support from CNPq under grant \#310309. LT thanks FAPESP for the support under grants \#2019/08341-8 and \#2022/02471-0. LC is grateful for support from the US National Science Foundation AS-1751874 and AS-2107070. KLP acknowledges funding from the UK Space Agency. The National Radio Astronomy Observatory is a facility of the National Science Foundation operated under cooperative agreement by Associated Universities, Inc.
We acknowledge the referee for the useful comments, which helped improved the model.

\section*{Data Availability}

The VLA data used in this paper are already published \citep{cho21}. The X-ray data are presented in Table \ref{tab:Table_1} of the present paper.



\bibliographystyle{mnras}
\bibliography{V5668Sgr} 




\appendix
\section{Relation between the physical parameter of the clumps}
\label{Ap_A}

\begin{figure*}
\includegraphics[width=17cm]{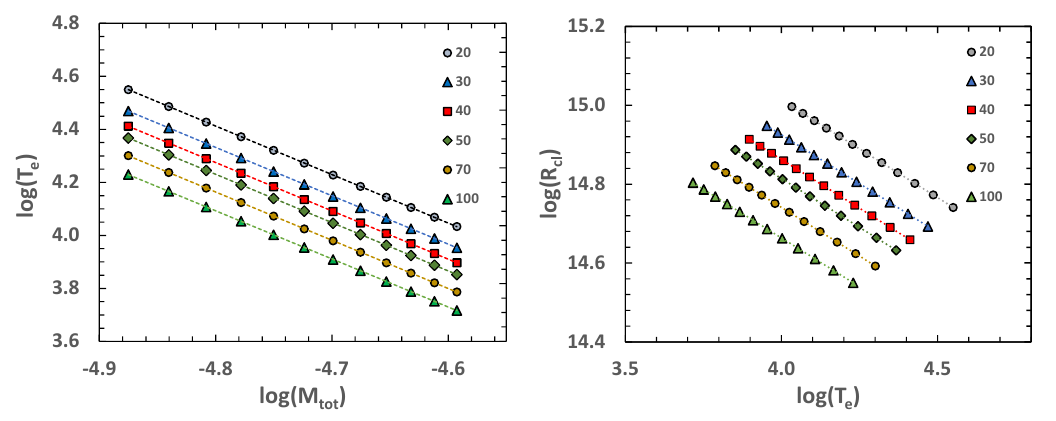}
	\caption{ Left panel: relationship between the electron temperature and the total mass of the clumps, for the models that fit the spectra of nova V5668 Sgr obtained with the VLA 966 days after eruption, for different number of clumps. Right panel: relationship between radius and temperature of the clumps, for the different numbers of clumps.}
    \label{fig:Fig_A1}
\end{figure*}

In Figure \ref{fig:Fig_A1} we present the relationship between temperature and radius of the clumps for the models that fit the spectra of nova V5668 Sgr obtained with the VLA 966 days after eruption, for different numbers of clumps: 20, 30, 40, 50, 70 and 100 (right panel), and  between the  temperature of the clumps and the total mass  of the ejecta, for the same number of clumps (left panel).

\begin{figure*}
\includegraphics[width=17cm]{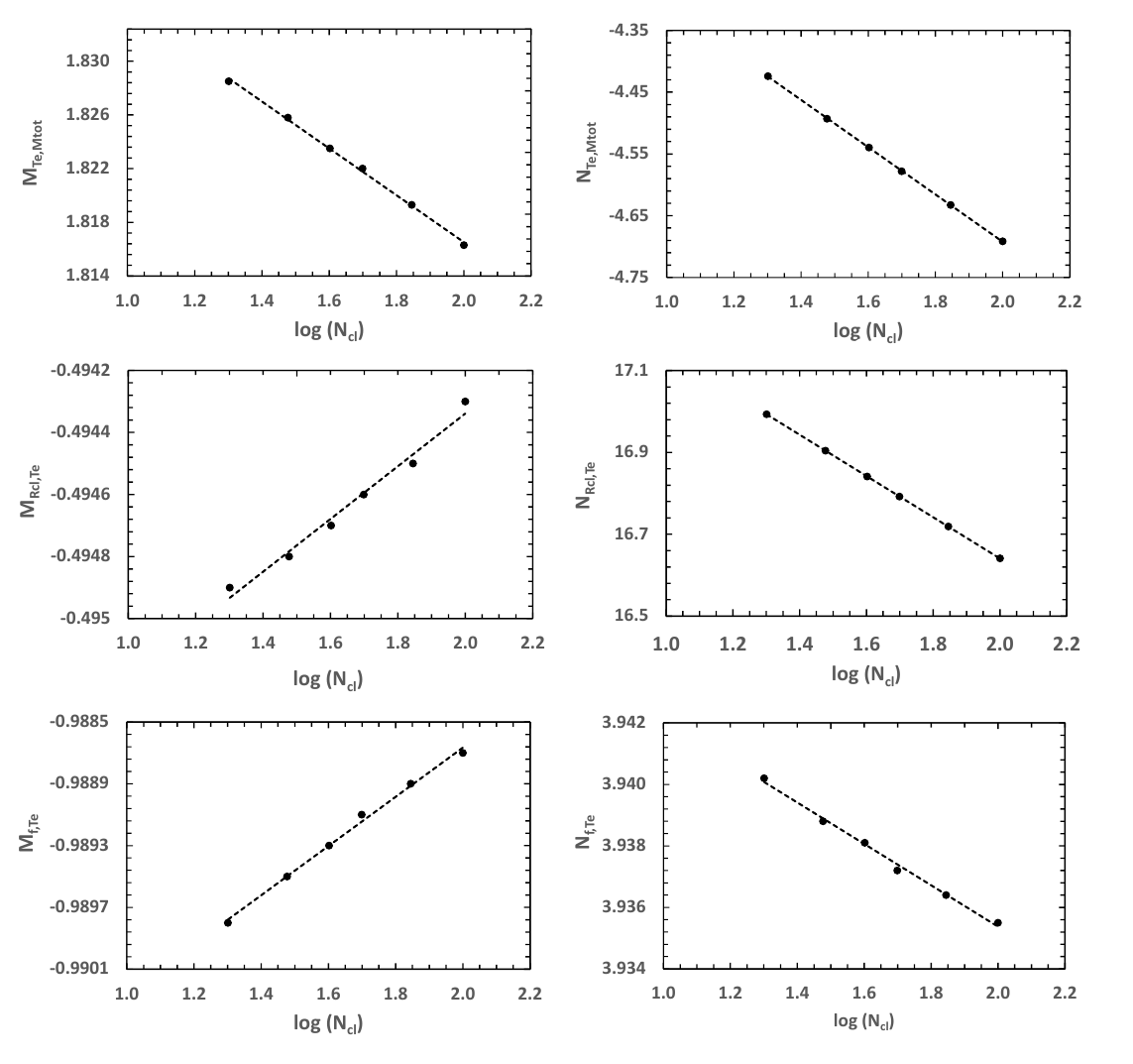}
	\caption{Data used to determine the coefficients $M_{\rm ij}$ and   $N_{\rm ij}$ as a function of the number of clumps $N_{\rm cl}$, defined in  equations \ref{Mij} and \ref{Nij}}
    \label{fig:Fig_A2}
\end{figure*}

In Figure \ref{fig:Fig_A2} we present the data used to determine the coefficients $M_{\rm ij}$ and   $N_{\rm ij}$ as a function of the number of clumps $N_{\rm cl}$, defined in   equations \ref{Mij} and \ref{Nij}, where i,j corresponds to the parameters $T_{\rm e}, M_{\rm tot}$ (top panel), $R_{\rm cl}, T_{\rm e}$ (middle panel) and $f, T_{\rm e}$ (bottom panel). 
\bsp	
\label{lastpage}
\end{document}